\documentclass[lettersize,journal]{IEEEtran}
\usepackage{amsmath,amsfonts}

\usepackage{algorithm}
\usepackage{algpseudocode}
\usepackage{array}
\usepackage{textcomp}
\usepackage{stfloats}
\usepackage{url}
\usepackage{verbatim}
\usepackage{graphicx}
\usepackage{cite}
\usepackage{xcolor}
\usepackage{subfigure}
\usepackage{booktabs}
\usepackage{multirow}
\usepackage{enumitem}  
\usepackage{cuted} 
\usepackage{array}
\usepackage{amssymb}
\usepackage{adjustbox}
\usepackage{booktabs}
\begin{document}

\title{A Geometry Map-Based Site-Specific Propagation Channel Model for Urban Scenarios}

\author{Junzhe Song,~\IEEEmembership{Student Member,~IEEE,}
Ruisi He,~\IEEEmembership{Senior Member,~IEEE,}
Mi Yang,~\IEEEmembership{Member,~IEEE,}\\
Zhengyu Zhang,~\IEEEmembership{Student Member,~IEEE,}
Shuaiqi Gao,
Xiaoying Zhang,
Bo Ai,~\IEEEmembership{Fellow,~IEEE,}

\thanks{
J. Song, R. He, M. Yang, Z. Zhang, S. Gao, and B. Ai are with the Schooof Electronics and Information Engineering, Beijing Jiaotong University, Beijing 100044, China (email: 25115063@bjtu.edu.cn; ruisi.he@bjtu.edu.cn; myang@bjtu.edu.cn; 21111040@bjtu.edu.cn; 25110046@bjtu.edu.cn; boai@bitu.edu.cn)

X. Zhang is with the College of Electronic Science and Technology, National University of Defense Technology, Changsha 410073, China (email: zhangxiaoying@nudt.edu.cn).


}}

\maketitle

\begin{abstract}
With the rapid deployments of 5G and 6G networks, accurate modeling of urban radio propagation has become critical for system design and network planning. However, conventional statistical or empirical models fail to fully capture the influence of detailed geometric features on site-specific channel variances in dense urban environments. In this paper, we propose a geometry map-based propagation channel model that directly extracts key parameters from a 3D geometry map and incorporates the Uniform Theory of Diffraction (UTD) to recursively compute multiple diffraction fields, thereby enabling accurate prediction of site-specific large-scale path loss and time-varying Doppler characteristics in urban scenarios. A well-designed identification algorithm is developed to efficiently detect buildings that significantly affect signal propagation. The proposed model is validated using urban measurement data, showing excellent agreement of path loss in both line-of-sight (LOS) and nonline-of-sight (NLOS) conditions. In particular, for NLOS scenarios with complex diffractions, it outperforms the 3GPP and simplified models, reducing the RMSE by 7.1 dB and 3.18 dB, respectively. Doppler analysis further demonstrates its accuracy in capturing time-varying propagation characteristics, confirming the scalability and generalization of the model in urban environments.
\end{abstract}


\IEEEdisplaynontitleabstractindextext

\IEEEpeerreviewmaketitle

\section{Introduction}
\IEEEPARstart{A}s 5G and 6G architectures continue to advance, ensuring reliable wireless connectivity in metropolitan areas has become a paramount priority \cite{5G,5G-1}. The rapid expansion of bandwidth-intensive frameworks, including smart city infrastructures \cite{V2X-1}, intelligent transportation systems \cite{V2X}, and vehicle-to-everything (V2X) networks \cite{he2019propagation,zhang2025channel}, requires highly reliable communication links within these challenging zones \cite{he2015characterization}. Unlike open environments, urban topographies are defined by densely packed high-rises and tightly constrained street grids. This distinct geometric layout yields a highly site-specific propagation channel \cite{Zemen-p1site}, where the exact physical arrangement of scatterers dictates signal degradation through severe reflections, blockages, and multi-edge diffractions \cite{zhang2025non}.

Fundamentally, radio wave propagation within city environments demonstrates several distinct traits \cite{canyon}: i) \textit{Severe multipath scattering:} The sheer volume of structural reflectors, including concrete facades and dense infrastructure, induces a proliferation of discrete multipath components (MPCs). Consequently, these areas suffer from substantially broader angular and delay spreads than unobstructed open spaces \cite{canyon-2}. ii) \textit{High degrees of non-stationarity and time variance:} The continuous mobility of transceivers (TX and RX) alongside transient environmental scatterers forces the propagation profile to fluctuate abruptly over short durations \cite{canyon-4}. iii) \textit{Asymmetric angular dispersion:} Confined street canyons typically restrict elevation spreads while massively inflating azimuthal spreads, with road intersections acting as critical junctions that inject complex supplementary scattering paths. iv) \textit{Profound spatial dependency:} The communication link is tightly coupled to the localized topology, meaning that building blueprints and street layouts dictate highly site-specific propagation behaviors. Given these complexities, it is imperative to develop robust channel modeling frameworks that not only trace the underlying physical propagation mechanisms but also meticulously integrate these localized geometric constraints \cite{canyon-5}.

Map-based channel models represent an important category of site-specific propagation models that integrate environmental geometry into channel characterization \cite{10742569}. Unlike stochastic models that rely on statistical fitting, map-based models utilize digital maps or 3D geometric databases to reconstruct the propagation environment, enabling more accurate modeling of path loss, MPCs, and spatial correlation. Representative standardization efforts, such as the 3GPP TR 38.901 hybrid map-based model \cite{zhu20213gpp}, incorporate map-based extensions to enhance realism in urban and indoor scenarios by combining deterministic ray-tracing with geometry-informed stochastic processes. Similarly, the METIS \cite{unknown} and WINNER II \cite{kyosti2007winner} frameworks adopt map-based modeling to support location-dependent channel generation and mobility-aware simulations. These developments highlight the increasing emphasis on geometry-aware modeling as wireless communication systems evolve toward higher frequencies and denser deployments. Despite these advances, most standardized map-based models still rely on deterministic simulation engines, where environmental information serves only as static input rather than as part of an inference process \cite{10839242}. Therefore, establishing a propagation model that directly extracts and utilizes environmental parameters from geometric maps and that derives channel characteristics based on physical propagation principles is essential. Such a model would provide a scalable and physically interpretable framework for accurately predicting radio channel behavior across diverse urban canyon environments \cite{he2024wireless}.

Extensive research has been dedicated to evaluating and modeling map-based propagation channels \cite{kyosti2017map,priebe2012calibrated,kanhere2023calibration,gan2018hybrid,pascual2018wireless,carton2016validation,10872967,wang2021ray,ying2025site,lim2017map,zhu2018novel,zhu2022map,grosse2016hybrid,alkhateeb2019deepmimo,alrabeiah2020deep,alkhateeb2023deepsense,zhao2023nerf2,bakirtzis2022deepray,hoydis2023sionna,10238401,cao2024raypronet,qiu2022pseudo,chatelier2024model,fu2025ckmdiff}. Generally, existing methodologies can be classified into three predominant paradigms: i) \textit{Deterministic modeling via geometric structures:} Using predominantly ray-tracing techniques, this category calculates MPCs through rigorous physical laws \cite{kanhere2024calibration}. Although they offer high precision, this fidelity is strictly dependent on the granular environmental databases in possession, including precise architectural layouts, street topologies, and the electromagnetic properties of materials \cite{10742569}. Consequently, this heavy data dependency severely restricts their deployment in massive-scale simulations, latency-sensitive applications, or regions lacking comprehensive geometric blueprints \cite{rainer2021scalable}. ii) \textit{Hybrid modeling approaches:} To strike a balance between computational burden and physical realism, these frameworks amalgamate deterministic tracing for primary propagation routes with stochastic processes for minor diffuse scattering \cite{zhu2022map}-\cite{grosse2016hybrid}. Despite their computational efficiency, these models are bottlenecked by the necessity of arduous parameter calibration. This reliance hampers their generalizability when empirical data is sparse and curtails their responsiveness to rapidly evolving, dynamic environments. iii) \textit{Data-driven approach:} These models use high-resolution maps or point clouds to train neural networks for predicting path loss, delay–angle spectra, or radio maps, showing strong potential for rapid inference. Physics-informed learning and domain adaptation further reduce the gap between simulated and measured data \cite{alkhateeb2019deepmimo,alrabeiah2020deep,alkhateeb2023deepsense,zhao2023nerf2,bakirtzis2022deepray,hoydis2023sionna,cao2024raypronet,qiu2022pseudo,chatelier2024model,fu2025ckmdiff,10238401}. However, their main limitations lie in the reliance on large amounts of labeled data, challenges in generalization, and the high retraining cost when the propagation environment undergoes significant changes.

In addition, several map-based models incorporate the Uniform Theory of Diffraction (UTD) or the Geometrical Theory of Diffraction (GTD) \cite{orfanidis2002electromagnetic}. Within the framework of geometrical optics, UTD/GTD introduce diffraction coefficients and asymptotic field solutions to accurately describe electromagnetic wave diffraction at geometric discontinuities such as building edges and corners, thereby maintaining path continuity even under nonline-of-sight (NLOS) conditions. This characteristic gives UTD/GTD inherent advantages in map-based modeling tasks. For instance, Du et al. derived a Fock-type integral-based “uniform correction term” that enables more physically consistent predictions near shadow boundaries, making UTD more suitable for use in map-based ray-tracing and preventing artificial interference artifacts \cite{du2023uniform}. Ortiz et al. developed an analytical/semi-analytical model using geometrically mapped urban blocks as input, combining GO and UTD to quantify diffraction and shadowing losses around street corners, and conducted an in-depth analysis of D2D link coverage \cite{brugarolas2023using}. Jarndal et al. further examined the diffraction loss approximation strategies adopted by map-based models such as METIS that employ GTD/UTD formulations \cite{jarndal2018mm}. 

However, when attempting to characterize multiple diffraction chains, most of these existing models remain deeply constrained by the traditional ray-tracing paradigm. Their core computational processes rely heavily on exhaustive path enumeration and recursive ray branching. In dense urban canyons with numerous building edges, this explicit tracking of individual ray trajectories leads to a prohibitive exponential explosion in computational complexity. Furthermore, existing deterministic engines typically treat massive 3D building databases as static, undifferentiated geometric inputs, lacking a dynamic mechanism to intelligently distill raw map data into the essential physical boundary conditions required by UTD. Consequently, there remains a critical methodological gap: achieving the rigorous physical fidelity of UTD-based multi-diffraction while fundamentally circumventing the computational bottleneck of traditional ray enumeration.
 
To bridge this gap, this paper proposes a novel geometry map-based propagation model that transitions the multi-diffraction computation from conventional path-enumeration to a stepwise recursive field transfer architecture. By intelligently extracting key geometric parameters from a 3D urban map, the model incorporates UTD to propagate the electric field layer-by-layer across successive scattering points. This methodology enables the accurate derivation of essential channel characteristics, such as path loss and Doppler distributions, without explicitly evaluating the full propagation integral for every single combinatorial path. The main contributions are summarized as follows:

\begin{itemize}
  \item We propose a geometry map-based propagation model that establishes precise UTD boundary conditions. Unlike conventional ray-tracing engines that suffer from exponential complexity due to ray branching, we derive a stepwise recursive equation for multiple diffractions, allowing the continuous, layer-by-layer transfer of electric fields between buildings.
  
  \item We develop a significant buildings identification algorithm that acts as a dynamic environment dimensionality reduction mechanism. It autonomously filters unstructured 3D map data to detect and extract only the dominant building topologies that dictate the signal propagation path, seamlessly linking raw geometric data with analytical UTD boundaries.
  
  \item We propose a significant buildings identification algorithm that automatically detects, within the geometric map, buildings that significantly affect the signal propagation path.

  \item We conduct urban channel measurements and model validation, demonstrating that the proposed model achieves excellent agreement with measured results under both LOS and NLOS conditions, and accurately reproduces Doppler characteristics, confirming its physical accuracy and generalization capability.
\end{itemize}
The remainder of this paper is organized as follows: Section II outlines the proposed propagation model. Section III presents the significant buildings identification. Section IV details the measurement and validation. Section V concludes the paper. 
\section{SYSTEM MODEL}
In this section, we propose a geometry map-based propagation model to characterize electromagnetic wave behavior in urban environments. The model extracts key geometric parameters from geometry maps. Based on the UTD, the model derives recursive formulations of multiple-diffraction electric fields and computes the corresponding path loss and doppler characteristics at the RX.
\subsection{Environmental geometric Parameterization}
\begin{figure}[t]
    \centering
    \includegraphics[width=\linewidth]{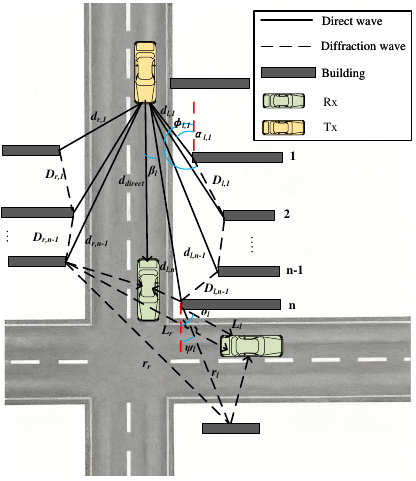}
    \caption{Diagram of signal propagation in an urban environment.}
    \label{fig:scene}
\end{figure}

In the proposed geometry map-based propagation model, the geometric parameters describing the urban environment are obtained directly from the geometry map database, enabling accurate characterization of each propagation segment determined by the geometry of TX, buildings, and RX, as illustrated in Fig. \ref{fig:scene}. The input parameters of the model are defined as follows:
\begin{itemize}
  \item \textbf{1) Distance vector between the TX and the buildings.} This parameter is used to calculate the free space loss, which is the primary factor in signal attenuation. This can be expressed as:
\begin{equation}
d=\left\{d_{l, 1}, d_{l, 2}, \cdots, d_{l, m}, d_{r, 1}, d_{r, 2}, \cdots, d_{r, n}\right\}
\end{equation}
  \item \textbf{2) Distance vector between adjacent buildings.} This parameter describes the propagation distance between successive segments in the process of multiple diffractions. This can be expressed as:
\begin{equation}
D=\left\{D_{l, 1}, D_{l, 2}, \cdots, D_{l, m}, D_{r, 1}, D_{r, 2}, \cdots, D_{r, n}\right\}
\end{equation}
  \item \textbf{3) Incident angle and wedge angle.} The incident angle $\alpha_i$ and wedge angle $\phi_i$ are defined as the angles between the incident wave direction and the normal of the wedge surface. They are used to determine the propagation region and to calculate the diffraction coefficients. These two angles dictate whether the signal is in the reflection, transmission, or diffraction region, thereby controlling the transition between different propagation mechanisms.
\end{itemize}
\subsection{Boundary Conditions}

\begin{figure}[t]
    \centering
    \includegraphics[width=\linewidth]{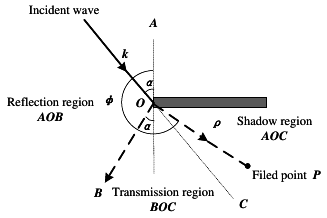}
    \caption{Propagation behavior of signals on building surfaces.}
    \label{fig:region}
\end{figure}

In urban electromagnetic propagation environments, building edges, corners, and surfaces cause electromagnetic waves to exhibit markedly different propagation characteristics in distinct regions, as illustrated in Fig. \ref{fig:region}. According to the UTD  \cite{orfanidis2002electromagnetic}, the propagation path at a wedge boundary can be divided into three physical regions. We assume $0 \le \alpha \le 90^\circ$ and distinguish three wedge regions defined by the half-plane and the directions corresponding to the reflected and transmitted rays:

\vspace{1em}
\begin{tabular}{ll}
reflection region $(AOB)$: & $0 \le \phi \le \alpha$ \\
transmission region $(BOC)$: & $\alpha \le \phi \le 2\pi - \alpha$ \\
shadow region $(COA)$: & $2\pi - \alpha \le \phi \le 2\pi$ \\
\end{tabular}
\vspace{1em}

In the case where $90^\circ \le \alpha \le 180^\circ$, the angle $\alpha$ is redefined such that it still lies within the range $0 \le \alpha \le 90^\circ$. The three wedge regions are now:

\vspace{1em}
\begin{tabular}{ll}
reflection region $(AOB)$: & $0 \le \phi \le \pi - \alpha$ \\
transmission region $(BOC)$: & $\pi -\alpha \le \phi \le \pi + \alpha$ \\
shadow region $(COA)$: & $\pi + \alpha \le \phi \le 2\pi$ \\
\end{tabular}
\vspace{1em}

According to \cite{orfanidis2002electromagnetic}, the following electromagnetic fields exist within the three regions:

\vspace{1em}
\begin{tabular}{ll}
reflection region: & $E_z = E_t + E_r +E_d$ \\
transmission region: & $E_z = E_t + E_d$ \\
shadow region: & $E_z = E_d$ \\
\end{tabular}
\vspace{1em}

We defined the incident, reflected, and diffracted fields:
\begin{equation}
\begin{gathered}
E_t=E_0 e^{j k D \cos \phi_i} \\
E_r=-E_0 e^{j k D \cos \phi_r} \\
E_d=-E_0 e^{-j k D} \frac{1-j}{2 \pi 2 \sqrt{\frac{k}{\pi D^{\frac{1}{2}}}}}\left(\frac{1}{\cos \left(\frac{\phi_i}{2}\right)}-\frac{1}{\cos \left(\frac{\phi_r}{2}\right)}\right)
\end{gathered}
\end{equation}
where $k = \frac{2\pi}{\lambda}$ is the wavenumber determined by the carrier wavelength $\lambda$, and $E_0$ is the incident electric field strength, and now:
\begin{equation}
\begin{aligned}
\phi_i &= \phi - \alpha \\
\phi_r &= \phi + \alpha
\end{aligned}
\end{equation}

\subsection{Multiple diffraction and recursive field calculations}
In a multi-diffraction propagation chain, the transmission of the electric field can be regarded as a stepwise recursive process. Suppose the signal sequentially passes through $n$ scattering points, such as building corners or edges; then the electric field received at the $i$-th scattering point consists of two components: (1) the composite field $E_z$ from the previous propagation stage and (2) the free-space attenuation component along the current propagation path, calculated based on geometric positions. The electric field can be expressed as:
\begin{equation}
E_{dir} =\frac{2 \sqrt{15 P_t}}{d} e^{-j k d}
\end{equation}
where $P_t$ denotes the transmitted power of the signal source.

To characterize multi-stage propagation effects, the electric field at each stage must be computed recursively, so that the field at any stage depends not only on the distance of the current propagation path, but also on the phase and amplitude resulting from the diffraction of previous stage. As shown in Fig. \ref{fig:scene}, the total electric fields at the first and second buildings can be expressed as follows:
\begin{equation}
\begin{gathered}
E_1=E_{dir}\left(d_1\right) \\
E_2=E_{dir}\left(d_2\right)+E_z\left[E_{1}, \phi_1, \alpha_1, D_1\right]
\end{gathered}
\end{equation}
The total electric field at level $i$ can be expressed recursively as follows:
\begin{equation}
E_n=E_{dir}\left(d_n\right)+E_z\left[E_{n-1}, \phi_{n-1}, \alpha_{n-1}, D_{n-1}\right]
\end{equation}

In the implementation, the initial condition for recursive propagation is the direct field $E_1$ at the first scattering point, after which $E_2, E_3, \dots, E_n$ are successively calculated through the transfer function at each stage. The key advantage of recursive computation is that it allows the electric field to be propagated layer-by-layer along multiple diffraction paths without explicitly evaluating the full propagation integral for each path. Moreover, each stage of propagation in the model directly corresponds to physically observable quantities, such as propagation distance, incident and outgoing angles, and building distribution, thereby providing clear physical interpretability.
\subsection{Total field calculation at the RX}
For the final stage of propagation, the electric field is influenced not only by the cumulative effects of multi-stage diffraction but also by the spatial propagation from the breakpoint to the RX. This process forms the final electric field $E_\mathrm{final}$, representing the field component generated by diffraction from the last scattering point to the RX, which constitutes the terminal field distribution of the entire propagation chain.

Based on the presence or absence of direct visibility between the TX and RX, propagation can be categorized into LOS and NLOS scenarios. In LOS scenarios, the RX simultaneously receives both the direct field from the TX and the diffracted field originating from the breakpoint. In NLOS scenarios, the direct path is blocked, so the RX can only capture the composite field resulting from multiple diffractions and reflections. Consequently, the total field in NLOS conditions includes additional reflection components and phase superposition effects. The two cases are discussed separately below.

In LOS scenarios, there exists a directly visible propagation path between the TX and RX. In addition to the direct free space field, diffracted components generated by building edges are also present. When the signal propagates from the TX through several diffraction points and undergoes a final diffraction at the breakpoint before reaching the RX, it forms the final electric field component $E_\mathrm{final}$. Therefore, the total electric field at the RX in an LOS scenario can be expressed as:
\begin{equation}
\begin{gathered}
E_{\mathrm{LOS}}=E_{\mathrm{dir}}+E_{\text {final }}^{(1)} \\
E_{\text {final }}^{(1)}=E_n D_s^{(I)} A_I e^{-j k L}
\end{gathered}
\end{equation}
where $E_{\text {final }}^{(1)}$ denotes the end winding field component after multiple windings \cite{zhang1997wide}.

Unlike LOS scenarios, the total electric field in NLOS conditions involves additional propagation mechanisms. After undergoing diffraction at the breakpoint, the signal also experiences reflections from buildings within the NLOS region. Consequently, the final electric field includes not only the first diffracted path but also composite components from reflected paths. The total electric field in an NLOS scenario can be expressed as:
\begin{equation}
\begin{gathered}
E_{\mathrm{NLOS}}=E_{\text {final }}+E_{\text {dir }} \\
E_{\text {final }}=E_n D_s^{(I)} A_I e^{-j k L}+R_{H, V} E_n D_s^{(I I)} A_{I I} e^{-j k r}
\end{gathered}
\end{equation}
where the first term of $E_{\text {final }}$ corresponds to the field component of the bypassed path, and the second term is the complex field reflected by the surface and then bypassed.

Where $A^{\mathrm{I}}$, $A^{\mathrm{II}}$ are amplitude coefficients, affected by the physical parameters from the breakpoint to the RX, indicating the influence of the path geometry, which can be expressed as:
\begin{equation}
\begin{gathered}
A^{\mathrm{I}}=\sqrt{\frac{d_n}{L\left(d_n+L\right)}} \\A^{\mathrm{II}}=\sqrt{\frac{d_n}{r\left(d_n+r\right)}}
\end{gathered}
\end{equation}

Let $\varepsilon_r$ be the relative permittivity of the reflecting building
wall. Thus, $R_{H}$ and $R_{v}$ can be calculated as
\begin{equation}
R_{H,V}=\frac{\cos \theta-a_{H, V} \sqrt{\varepsilon_r-\sin ^2 \theta}}{\cos \theta+a_{H, V} \sqrt{\varepsilon_r-\sin ^2 \theta}}
\end{equation}
where $a_{H} = 1$ and $a_{H} = 1/\varepsilon_r$ correspond to $R_{H}$ and $R_{v}$, respectively. As $\varepsilon_r \to \infty$, implying a perfectly conducting wall, one obtains $R_{H} = -1$ and $R_{V} = 1$. Results published in the literature on mobile radio communications indicate that the value of $\varepsilon_r$ makes negligible differences to the propagation measurements and calculations \cite{zhang1997wide}.

$D_s^{(I)}$, $D_s^{(II)}$ are the diffraction coefficients corresponding to the two paths, respectively, calculated based on the modified Fresnel integral:
\begin{equation}
\begin{aligned}
D_s^{(I I)}=& \frac{-e^{-j \pi / 4}}{2 \sqrt{2 \pi k}} \\
& \left(\frac{F\left(X_1\right)}{-\sin ((\theta-\beta) / 2)} - \frac{F\left(X_2\right)}{-\cos ((\theta+\beta) / 2)}\right)
\end{aligned}
\end{equation}

Moreover, X1 and X2 are defined by
\begin{equation}
\begin{aligned}
\sqrt{X_1} & =\sqrt{2 k L^{II}}|\sin [(\theta-\beta) / 2]| \\
\sqrt{X_2} & =\sqrt{2 k L^{II}} \mid \cos [(\theta+\beta) / 2] \\
L^{II} & =\frac{r}{1+r / d_n} .
\end{aligned}
\end{equation}

The expression of $D_s^{(I)}$ follows the same form as $D_s^{(II)}$, 
with the substitution of $\theta \rightarrow \psi$ and $L^{II} \rightarrow L^{I} =\frac{L}{1+L / d_n}$.

Furthermore, $F(X)$ is the transition function defined in  \cite{zhang1997wide}. In the actual computation, it is appropriate to write for $F(X)\approx\sqrt{\pi X} e^{j(\pi / 4+X)}$. In other cases, it is advantageous to write as follows:
\begin{equation}
\begin{aligned}
F(X)= & \sqrt{\pi X} e^{j(\pi / 4+X)} \\
& -2 j \sqrt{X} e^{j X} \int_0^{\sqrt{X}} e^{-j \tau^2} d \tau
\end{aligned}
\end{equation}

\subsection{Path Loss Calculation}
\begin{figure}[t]
    \centering
    \includegraphics[width=\linewidth]{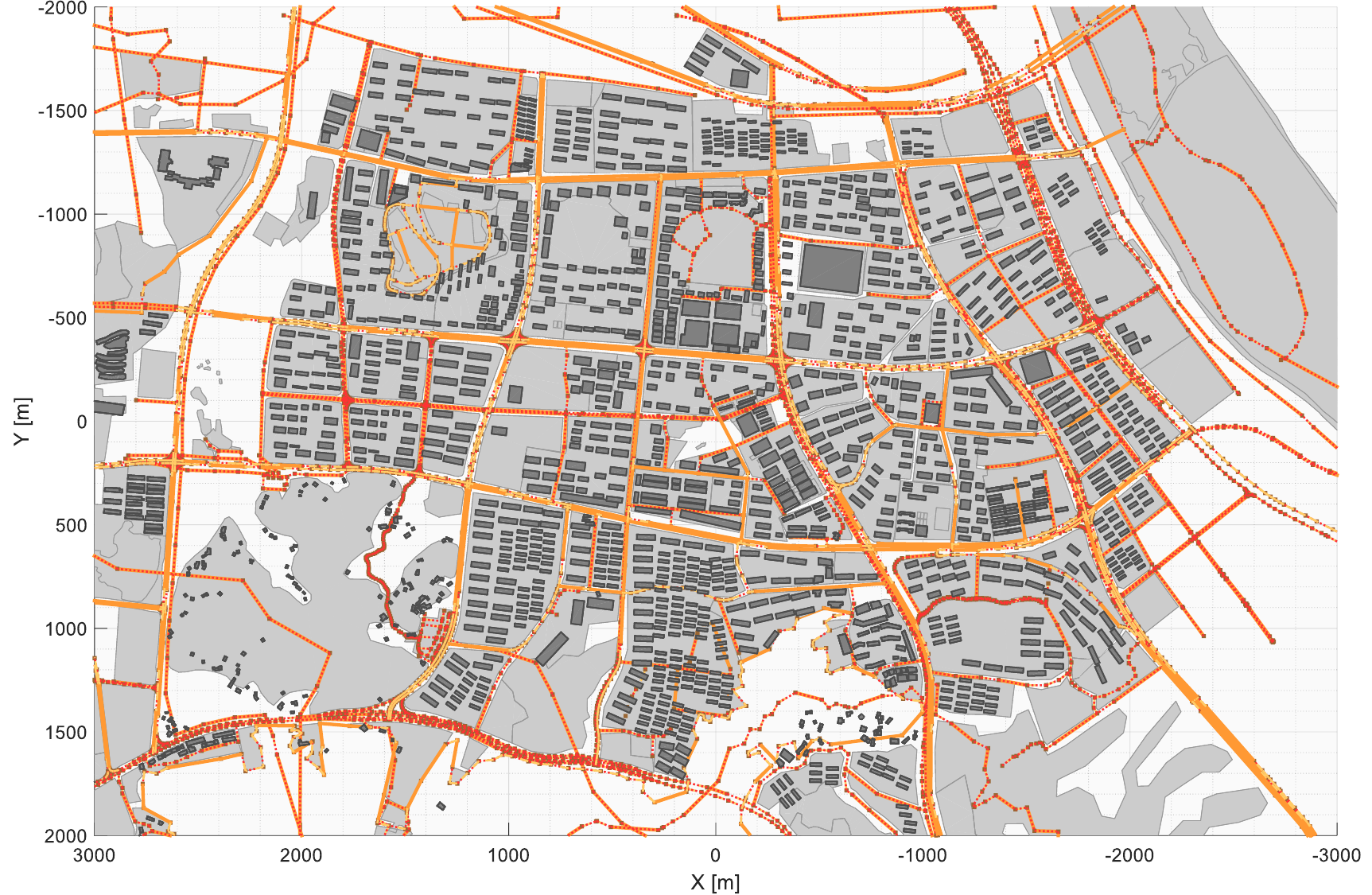}
    \caption{3D geometric map.}
    \label{fig:map}
\end{figure}
After multiple diffractions and recursive field superposition, the resulting electric field strength must be converted into received power to compute the path loss. The received power is determined by the square of the electric field strength and its relationship with the intrinsic impedance of free space. According to the fundamental principles of electromagnetics, the power density per unit area can be expressed as:
\begin{equation}
S = \frac{|E_i|^2}{2 \eta_0}, \quad i \in \{\text{LOS}, \text{NLOS}\}
\end{equation}
where $\eta_0=120\pi$ is the intrinsic impedance of the free space. Under far-field conditions, the effective aperture of the receiving antenna is related to wavelength $\lambda$ and antenna gain $G_r$ as follows:
\begin{equation}
A_e=\frac{G_r \lambda^2}{4 \pi}
\end{equation}
Thus, the received power can be expressed as:
\begin{equation}
P_r=\frac{\lambda^2 G_r\left|E_{\text {i }}\right|^2}{8 \pi \eta_0}
\end{equation}
Finally, the total path loss can be expressed as:
\begin{equation}
P L_{d B}=-10 \log _{10}\left(\frac{P_r}{P_t}\right)
\end{equation}

Through the above computational procedure, the model establishes a mapping from environmental geometry parameters to path loss. By inputting geometry parameters, the model recursively computes the electric field strength at the RX and ultimately determines the path loss. For different building layouts and street widths, variations in propagation distance and wedge angle among input parameters lead to differences in diffraction coefficients. By dynamically updating these input quantities, the model enables continuous prediction of path loss in complex urban environments.
\subsection{Doppler Characteristics}
In time-varying wireless channels, the Doppler effect serves as a key feature characterizing the dynamic behavior of the channel. It can be obtained by simulating the RX motion along a road trajectory. Specifically, at each sampling position, the model calculates the angle of arrival and the path power of all propagation paths reaching the RX, from which the corresponding Doppler shifts are derived. The collection of these Doppler shifts over spatial positions, together with their statistical measure—the RMS Doppler spread—provides a quantitative description of the Doppler characteristics.

For a single propagation path, the Doppler shift is determined by the projection of the RX velocity onto the direction of wave propagation, expressed as:
\begin{equation}
f_{d, i}=\frac{v \cdot \hat{u}_i}{\lambda}
\end{equation}
where $v$ is the RX velocity vector and $\hat{u}_i$ denotes the unit vector of the incident path direction. At a given RX position, the model aggregates the Doppler shifts of all effective paths $\left\{f_{d, 1}, f_{d, 2}, \ldots, f_{d, M}\right\}$, using the received path power $P_i$ as a weighting factor to compute the power-weighted RMS Doppler spread, which can be expressed as:
\begin{equation}
\sigma_{\mathrm{d}}=\sqrt{\frac{\sum_i P_i\left(f_{d, i}-\bar{f}_{d, \mathrm{w}}\right)^2}{\sum_i P_i}}, \quad \bar{f}_{d, \mathrm{w}}=\frac{\sum_i P_i f_{d, i}}{\sum_i P_i}
\end{equation}

This formulation captures the distinct contributions of different propagation paths to the overall Doppler spectrum. By coupling geometric features from map data with the previously derived recursive electric field computation model, the framework enables the generation of spatially continuous Doppler field distributions across the environment.
\begin{table}[t]
\centering
\caption{Notation for Significant Buildings Identification}
\begin{tabular}{l p{6cm}}
\toprule
\textbf{Symbol} & \textbf{Definition} \\
\midrule
$TX$ & Transmitter position \\
$RX, \; r$ & Receiver position (candidate point along route) \\
$bp$ & Breakpoint (intersection point for NLOS case) \\
$\mathcal{B}$ & Set of all buildings in the map \\
$\mathcal{B}_{side}$ & Candidate buildings located on each side of TX--RX (Algorithm 1 output) \\
$\mathcal{B}_{vis}$ & Final set of visible buildings after filtering \\
$V$ & Vertex set of buildings \\
$f_{\text{block}}(\cdot)$ & Occlusion test function: check if line segment is blocked by building faces \\
$f_{\text{proj}}(\cdot)$ & Projection function: compute orthogonal projection of a point onto a line \\
$f_{\text{side}}(\cdot)$ & Side classification function: determine whether a point lies on the left/right side of a line \\
LOS flag & Boolean indicator whether TX--RX is LOS \\
$\mathcal{B}_{left}$ & Buildings on the left side of bp--RX (NLOS case only) \\
\bottomrule
\end{tabular}
\end{table}
\section{Significant Buildings Identification}
The 3D geometric map serves as the source of physical parameters for the model input, as shown in Fig. \ref{fig:map}. Based on this map information, we propose a significant buildings identification algorithm designed to automatically detect buildings that have a significant impact on the signal propagation path. The algorithm consists of two core modules: initial building identification and visible building identification. The former determines the set of buildings located on both sides of the propagation path that may potentially act as scatterers, while the latter refines this set by applying occlusion detection and projection-based geometric analysis to eliminate obstructed buildings, thereby obtaining the final set of significant buildings.

\begin{algorithm}[t]
\caption{Initial Building Identification}
\begin{algorithmic}[1]
\Require TX, RX route, $V$, $F$, $G$
\Ensure $\mathcal{B}_{side}$: candidate buildings classified by side

\State Initialize $\mathcal{B}_{side} \gets \emptyset$

\For{each RX position $r$ in RX route}
    \State Compute vector $\vec{d} \gets r - TX$
    \State Determine LOS/NLOS:
    \If{$\neg \bigcup_{f \in F} f_{\text{block}}(TX, r, f)$}
        \State LOS $\gets$ True
    \Else
        \State LOS $\gets$ False
    \EndIf
    
    \If{LOS}
        \For{each building group $g \in G$}
            \State Select vertices $p \in V_g$ within TX--RX segment
            \State $p_{proj} \gets f_{\text{proj}}(p, TX\!-\!r)$
            \State Retain top-level vertices
            \For{each vertex $p$}
                \State $\text{side}(p) \gets f_{\text{side}}(p, TX\!-\!r)$
            \EndFor
            \State Add $g$ to $\mathcal{B}_{side}$ according to vertex sides
        \EndFor
    \Else
        \State Compute breakpoint $bp$ along RX direction
        \State Repeat LOS procedure for TX--bp vector
        \State For bp--RX vector, consider only left-side buildings
        \State Append results to $\mathcal{B}_{side}$
    \EndIf
\EndFor

\State \Return $\mathcal{B}_{side}$
\end{algorithmic}
\end{algorithm}
\subsection{Map and Parameter Setting}
The 3D geometric map is composed of a set of vertex $V$, a set of faces $F$, and a set of building groups $G$. Each building consists of multiple polygonal surfaces, with vertex coordinates stored in either geographic or local coordinate systems. The positions of the TX and RX are defined by measurement trajectories or simulation paths, and the system analyzes the propagation geometry point by point along the RX movement trajectory.

For clarity of the algorithmic description, Table I lists the main symbols and their corresponding meanings used in the building identification process. In practical computation, the system first determines the straight-line direction between the TX and RX as the reference propagation vector, based on which it identifies potential buildings located within the propagation corridor. When occlusion occurs, the algorithm automatically calculates the breakpoint (bp) along the NLOS path and performs identification and filtering separately within the two subsegments, TX–bp and bp–RX. This ensures the feasibility of the model and physical consistency under NLOS conditions. The detailed mathematical definitions of the auxiliary functions used in the identification process are provided in the Appendix.
\begin{algorithm}[t]
\caption{Visible Building Identification}
\begin{algorithmic}[1]
\Require $\mathcal{B}_{side}$, $V$, TX, $r$, LOS flag for current RX, $bp$ if NLOS
\Ensure $\mathcal{B}_{vis}$: set of visible buildings

\State Initialize $\mathcal{B}_{vis} \gets \emptyset$

\If{LOS}
    \State $\mathcal{B}_{check} \gets \mathcal{B}_{side}$ 
\Else
    \State $\mathcal{B}_{check} \gets \mathcal{B}_{side}$ along TX--bp
    \State $\mathcal{B}_{left} \gets$ left-side buildings along bp--RX
    \State Append $\mathcal{B}_{left}$ to $\mathcal{B}_{check}$
\EndIf

\For{each side in \{left, right\}}
    \State Sort $\mathcal{B}_{check}[side]$ by perpendicular distance to relevant line
    \State occlusionFlag $\gets$ False
    \For{each building $B$ in sorted order}
        \State isVisible $\gets$ True
        \For{each vertex $v \in B$}
            \State $v_{proj} \gets f_{\text{proj}}(v, \text{relevant line})$
            \For{each previously accepted building $B'$ in $\mathcal{B}_{vis}$}
                \If{$f_{\text{block}}(v, v_{proj}, V_{B'})$}
                    \State isVisible $\gets$ False
                    \State occlusionFlag $\gets$ True
                    \State \textbf{break}
                \EndIf
            \EndFor
            \If{not isVisible} \textbf{break} \EndIf
        \EndFor
        \If{isVisible}
            \State Add $B$ to $\mathcal{B}_{vis}$
        \EndIf
    \EndFor
\EndFor

\State \Return $\mathcal{B}_{vis}$
\end{algorithmic}
\end{algorithm}

\subsection{Initial Building Identification}
The initial building identification algorithm aims to select candidate scatterers from the global building set that potentially affect the propagation path. The core idea is to identify buildings located near the TX–RX propagation path through geometric projection and occlusion analysis and to classify them according to their position on either side of the path, providing input for subsequent visibility determination.

The algorithm inputs include the TX, RX trajectory, $V$, $F$, and $G$. The output is the set of classified candidate building $\mathcal{B}_\text{side}$, divided into left and right groups.
The main steps are as follows:

First, for each position point $r$ along the RX trajectory, the vector $\mathbf{d} = r - \text{TX}$ is calculated to determine the propagation direction from the transmitter to the receiver. Next, occlusion detection is performed along this propagation line to determine whether any building face $F_f \in F$ blocks the line of sight between TX and $r$. If no occlusion exists, the location is defined as a LOS scenario; otherwise, it is identified as a NLOS scenario.

In the LOS case, the algorithm traverses all building groups $G_g \in G$ and selects those whose vertex sets are located near the TX–RX line segment. Then, through the projection function $f_{\text{proj}}(\cdot)$, the projection coordinates of each vertex along the propagation direction are calculated to exclude low-level or rear-side vertices outside the effective propagation range, retaining only the top or boundary vertices. For retained vertices, the function $f_{\text{side}}(\cdot)$ is invoked to determine their lateral position relative to the propagation direction, thus completing the left–right-hand classification of scatterers. Finally, the buildings are assigned to the set $\mathcal{B}_\text{side}$ according to their side classification.

In NLOS scenarios, the algorithm first computes the breakpoint $bp$, defined as the intersection point between the TX–RX direction and the obstructing building, which divides the TX–RX line into two propagation segments: TX–bp and bp–RX. The TX–bp segment repeats the building filtering process used in the LOS case to identify edge buildings responsible for initial diffraction, while the bp–RX segment considers only buildings on the left side, which may contribute to secondary scattering after diffraction. The results from the two segments are then combined to form the complete set of candidates $\mathcal{B}_\text{side}$.

\subsection{Visible Building Identification}
\begin{figure}[t]
    \centering
    \includegraphics[width=\linewidth]{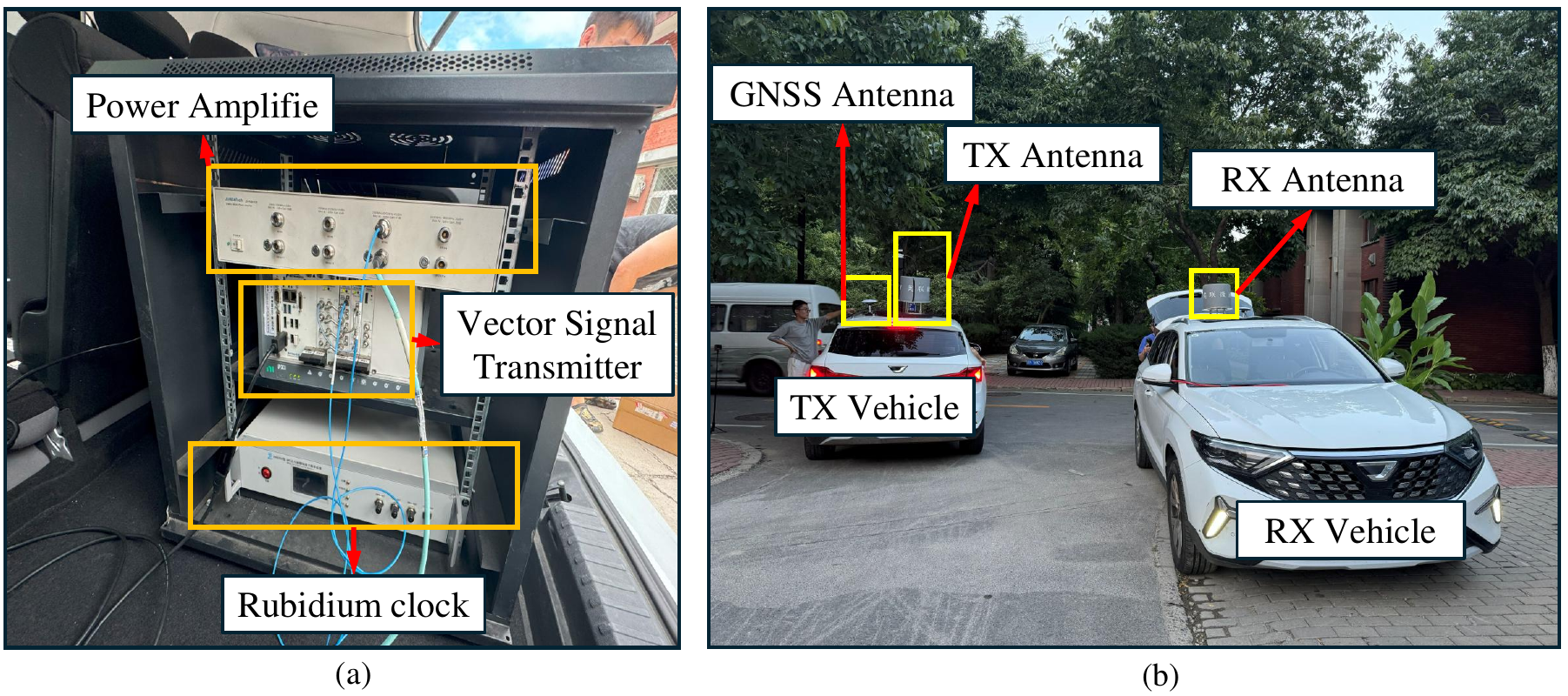}
    \caption{Measurement system architecture and key equipment: (a) TX vehicle equipment, (b) measurement vehicles and antennas.}
    \label{fig:measurement}
\end{figure}
After obtaining the set of candidate buildings $\mathcal{B}_{\text{side}}$, it is still necessary to determine which buildings actually participate in signal propagation. Based on this, the visible building identification algorithm introduces occlusion detection and geometric projection analysis to construct a visibility filtering process, thereby deriving the final set of visible buildings $\mathcal{B}_{\text{vis}}$.

The algorithm takes as input the candidate building set $\mathcal{B}_{\text{side}}$, the vertex set $V$, the transmitter $TX$, the receiver position $r$, the LOS indicator, and the breakpoint $bp$ for the NLOS conditions. The output is the visible building set $\mathcal{B}_{\text{vis}}$ corresponding to the current receiver position. The main steps are as follows:

First, the algorithm determines the buildings to be examined according to the LOS or NLOS condition. In the LOS scenario, the checking set $\mathcal{B}_{\text{check}}$ is equal to $\mathcal{B}_{\text{side}}$. In the NLOS case, $\mathcal{B}_{\text{check}}$ is composed of candidate buildings along the $TX$--$bp$ path and the buildings located on the left side of the $bp$--$RX$ path. This step ensures that all potential propagation paths are covered even under obstructed conditions.
\begin{figure}[t]
    \centering
    \includegraphics[width=\linewidth]{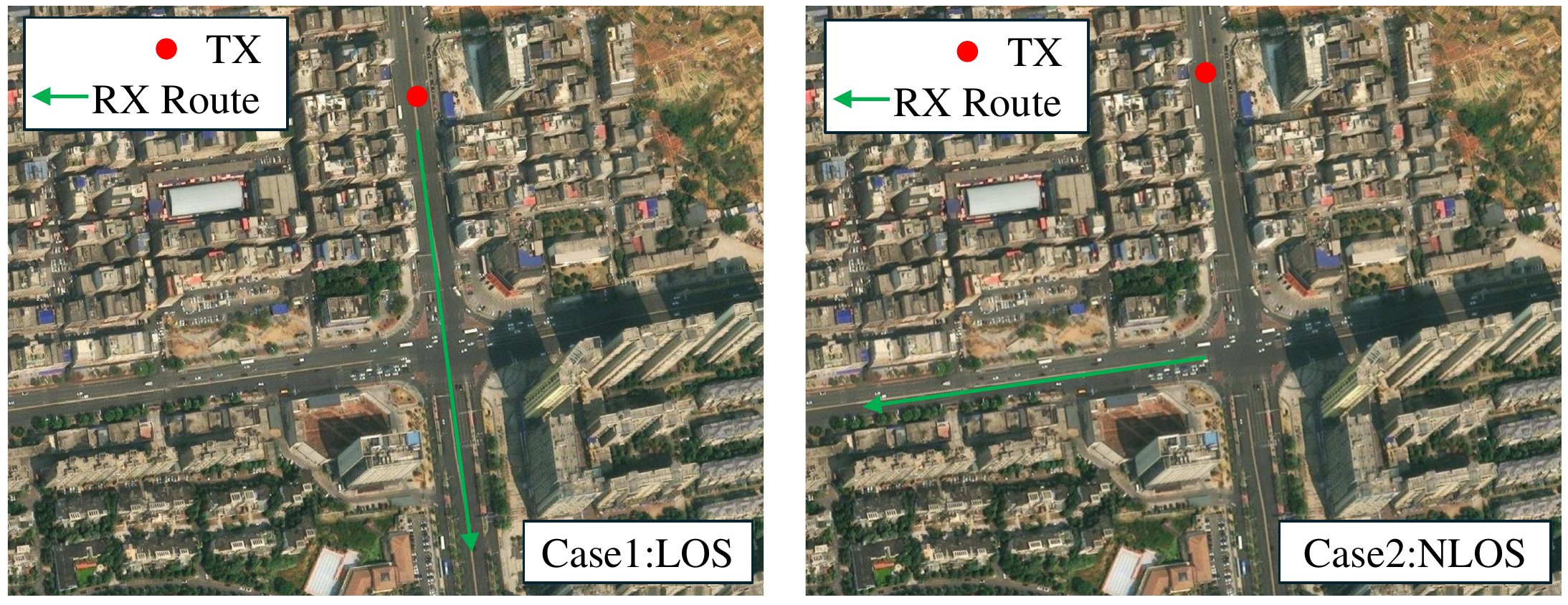}
    \caption{Measurement Scenarios.}
    \label{fig:route}
\end{figure}

The algorithm then processes the building sets on both sides of the propagation path independently. For each side, the buildings are sorted according to their perpendicular distance to the propagation baseline, and those closer to the baseline are given higher priority for visibility evaluation. For each building $B$, the projection of each vertex is computed as \[v_{\text{proj}} = f_{\text{proj}}(v, \text{line}),\] and occlusion detection is performed sequentially with the surfaces of buildings already identified as visible in $\mathcal{B}_{\text{vis}}$. If the line segment connecting a vertex to its projection is blocked by the set of surfaces $V_{B'}$ of any building $B' \in \mathcal{B}_{\text{vis}}$, the building $B$ is determined to be occluded and excluded from propagation. Otherwise, it is marked as visible and added to $\mathcal{B}_{\text{vis}}$.

The core of the algorithm lies in its hierarchical visibility filtering mechanism. By performing detection and cumulative occlusion assessment in a near-to-far sequence, the algorithm progressively constructs a layered visibility structure, effectively avoiding redundant calculations and unnecessary geometric traversals. Moreover, since the algorithm operates independently at each position $RX$, it can continuously compute visible scatterer variations along dynamic receiver trajectories, enabling real-time updates of the propagation environment with respect to receiver movement.

The resulting set of $\mathcal{B}_{\text{vis}}$ constitutes the effective scatterer ensemble in the current propagation geometry. This set not only reflects the geometric distribution characteristics of urban buildings, but also identifies the key structural elements that influence the propagation of the electromagnetic wave. By integrating this with the physical propagation model proposed in the previous section, the system can visualize real-time changes in building visibility and their effects on path loss, thereby achieving a complete link from geometric modeling to propagation prediction.

\section{Validation and Evaluation}
We validate and evaluate the proposed propagation model using measured channel data. Comparative analyzes are conducted on two test routes in typical urban canyon environments: one representing LOS conditions and the other NLOS conditions, to assess the applicability of model and prediction accuracy under different propagation scenarios.
\begin{figure*}[t]
    \centering
    \includegraphics[width=\linewidth]{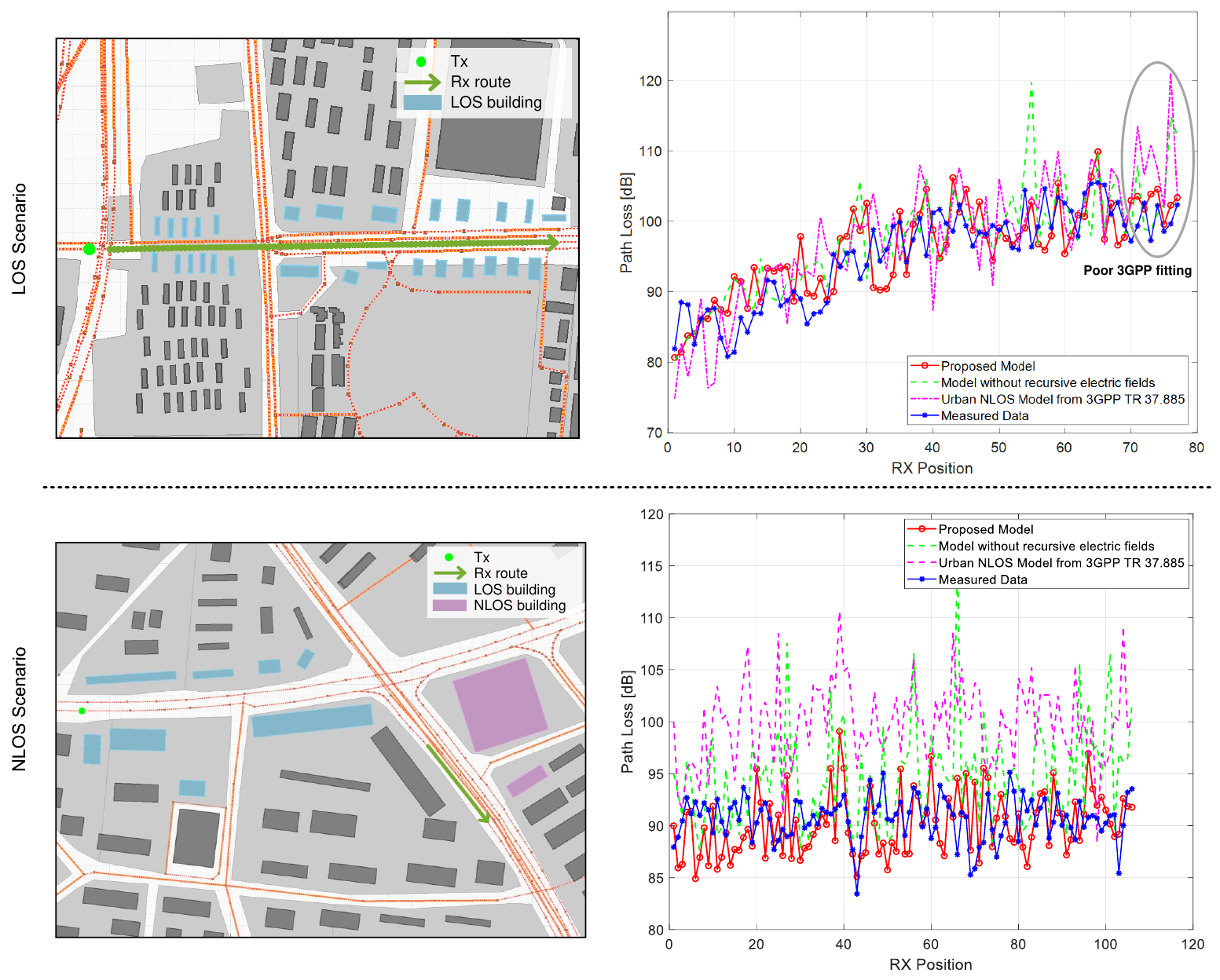}
    \caption{Path loss comparison between the proposed model, other models, and measured data.}
    \label{fig:route}
\end{figure*}
\subsection{Measurement Campaign}
We conducted measurement campaigns using the equipment shown in Fig. \ref{fig:measurement}. The measurement system consists of the TX and RX subsystems integrated into vehicles, with the core components being a vector signal generator (VSG) and a vector signal analyzer (VSA). Specifically, the VSG and VSA are NI PXIe-5673 and NI PXIe-5663, respectively. The sounding signal is a broadband multi-carrier signal with 513 subcarriers over a bandwidth of 30 MHz, centered at 5.8 GHz.

For precise time synchronization, two rubidium clocks disciplined by Global Navigation Satellite System (GNSS) signals are employed. These clocks also provide real-time longitude and latitude coordinates, enabling accurate positioning of both the TX and RX. The measurement bandwidth is 30 MHz, which results in a delay resolution of 33.3 ns. This delay resolution means that only MPCs with propagation distance differences exceeding 10 m can be distinguished in the delay domain. For vehicular communications with sub-6 GHz, the available bandwidth is generally 20–30 MHz, which is similar to the measurement configuration in this article. The acquisition rate of channel snapshots is 120 snapshots/s when measured.

The measurements were conducted in Changsha, China. The measurement area covers major roads within a 4 km × 3 km urban district in Fig. \ref{fig:map}. The streets under measurement are flanked by densely packed buildings reaching several tens of meters in height, forming a typical urban canyon environment. As illustrated in Fig. \ref{fig:route}, the measurement routes comprise two cases:
\begin{itemize}
  \item \textbf{Case1: LOS scenarios.} The TX and RX vehicles were on the same road, with the RX vehicle positioned ahead of the TX vehicle. The TX vehicle was parked at the roadside, while the RX vehicle maintained an average speed of 20 km/h to ensure consistency in data collection.
  
  \item \textbf{Case2: NLOS scenarios.} The RX vehicle traveled on a road adjacent to and approximately perpendicular to the one where the TX was located. TX vehicles parked on the side of the road at the beginning of the process, and the RX maintained the same speed as in the LOS cases.
\end{itemize}

Measurements were conducted in the absence of other nearby vehicles to ensure that additional vehicular movements did not influence the results used for analysis.
\begin{figure}[t]
\centering
\subfigure[]{\includegraphics[width=3.3in]{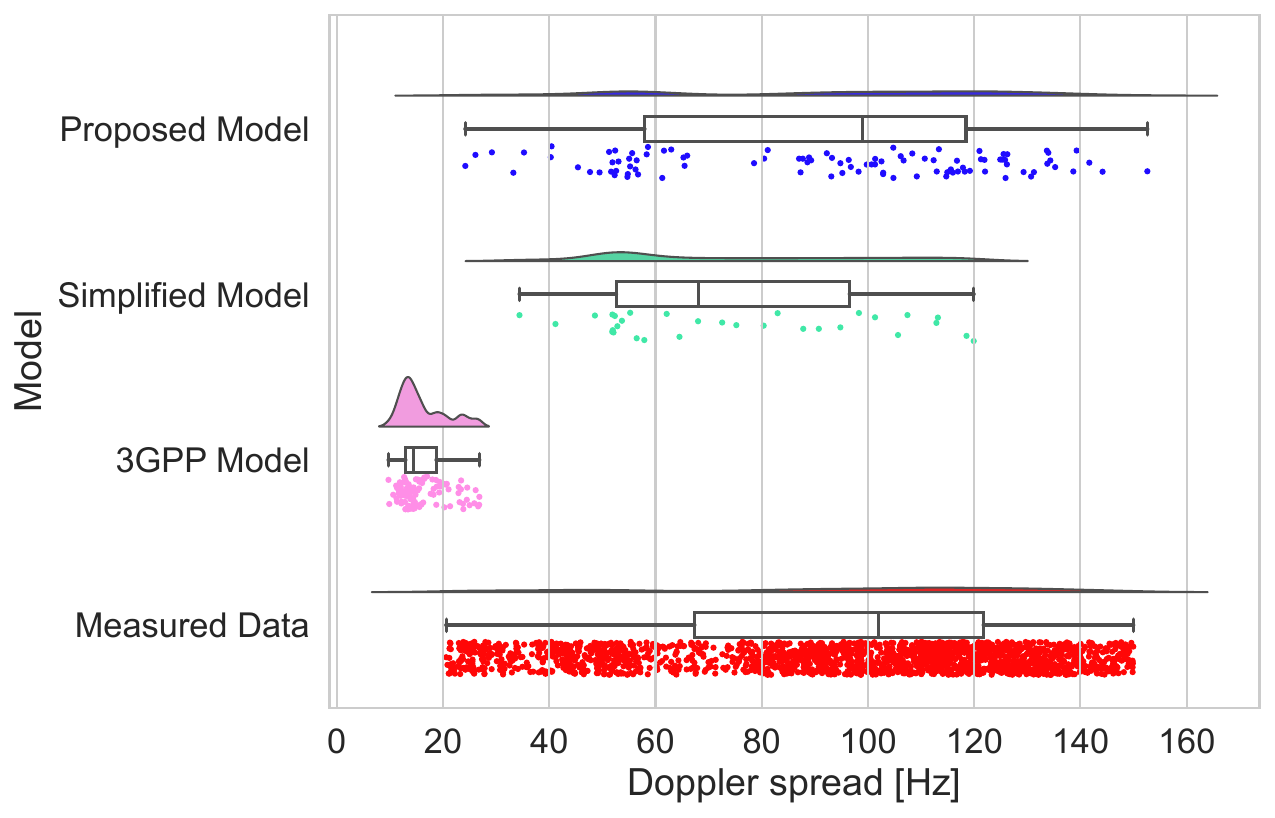}}
\subfigure[]{\includegraphics[width=3.3in]{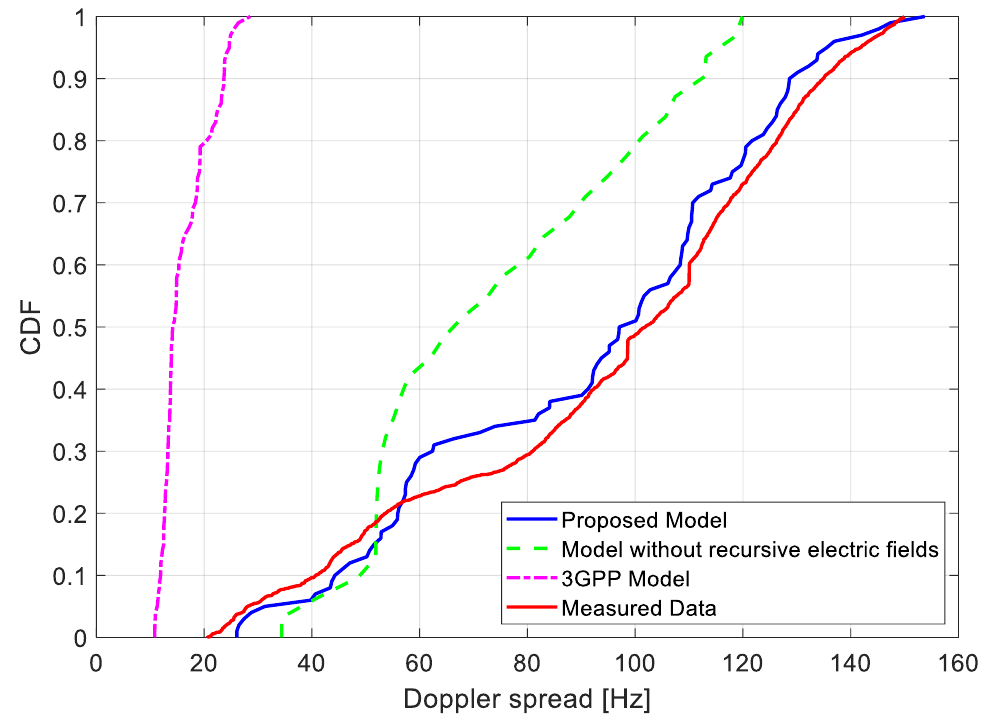}}
\caption{Doppler Spread Analysis in the LOS Route. (a) Scatter density map. (b) CDF comparison.}
\label{fig:LOS}
\end{figure}
\subsection{Path Loss validation}
Two test routes are selected from typical urban environments for analysis: one under LOS conditions and the other under NLOS conditions, as shown in Fig. \ref{fig:route}. The figure also illustrates the effectiveness of the significant building identification process, where buildings located along both sides of the main propagation path and not blocked by obstacles are accurately detected and highlighted. During model validation, both the measured path loss and the predicted results from three models were calculated for each route: (1) the proposed propagation model, (2) the 3GPP TR 37.885 standard path loss model, and (3) a simplified model that neglects multiple diffraction effects (considering only the propagation from the last diffraction point to the receiver). The root-mean-square error (RMSE) was adopted as the evaluation metric to quantify the deviation between the model predictions and the measured data, as shown in Table II.
\begin{figure}[t]
\centering
\subfigure[]{\includegraphics[width=3.3in]{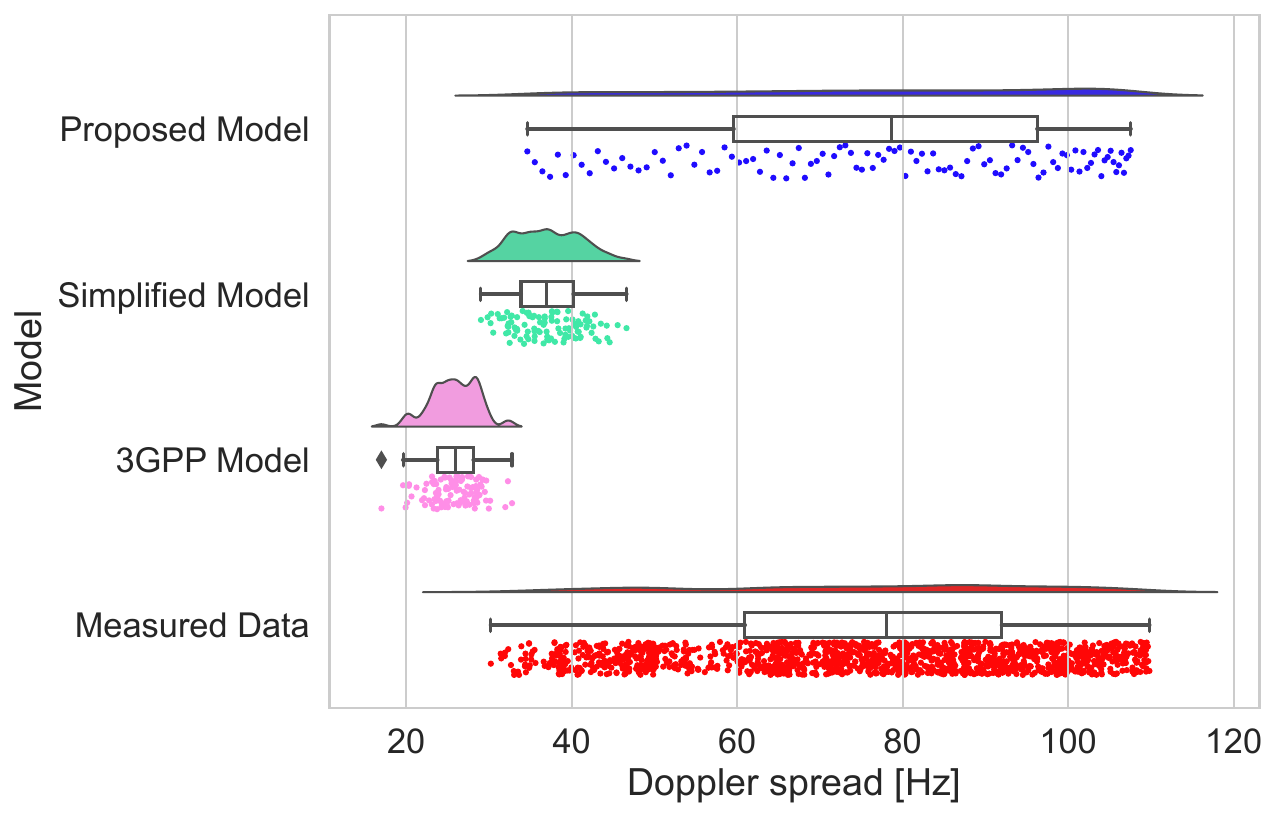}}
\subfigure[]{\includegraphics[width=3.3in]{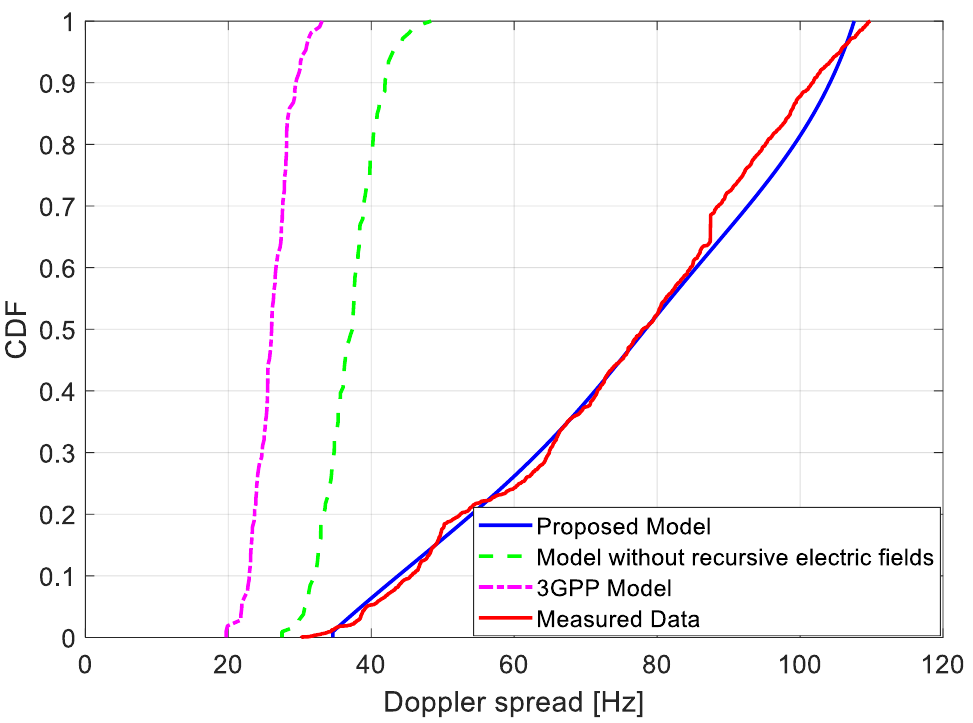}}
\caption{Doppler Spread Analysis in the NLOS Route. (a) Scatter density map. (b) CDF comparison.}
\label{fig:NLOS}
\end{figure}

In the LOS route, signal propagation is primarily dominated by the direct free-space component. As shown in Fig. \ref{fig:route}, all three models capture the overall trend of path loss variation, but noticeable differences appear in local fitting accuracy. The proposed model closely follows the measured results along the entire route, accurately reproducing small-scale fluctuations caused by weak reflections and scattering from surrounding buildings. In contrast, the simplified model and the 3GPP model exhibit evident deviations in certain regions—particularly where local reflections or partial obstructions occur. The simplified model neglects these secondary propagation effects, while the 3GPP model, being purely empirical, fails to account for the geometric characteristics of the actual environment. As a result, both tend to either overestimate or underestimate the path loss in specific segments, leading to reduced local prediction accuracy.

In contrast, the NLOS route presents a much more complex propagation environment. In this case, buildings completely block the line of sight, and the signal relies mainly on diffraction and multiple diffraction effects along building edges. As shown in Fig. \ref{fig:route}, the proposed model exhibits a significantly better prediction performance in the NLOS scenario. Specifically, the proposed model achieves an RMSE of 3.59 dB, with the predicted curve closely matching the measured path loss trend, accurately capturing both the abrupt attenuation transitions and local energy enhancement regions. The simplified model yields an RMSE of 6.77 dB, showing larger errors and fluctuations, making it difficult to capture the detailed variations in path loss—further confirming the importance of multiple diffraction mechanisms in urban propagation modeling. The 3GPP standard model exhibits the highest RMSE of 10.69 dB, with generally overestimated results, indicating that traditional empirical models fail to accurately represent the diffraction characteristics caused by building obstructions in NLOS environments.

In general, the proposed propagation model demonstrates stable predictive capability in both LOS and NLOS scenarios, with particularly notable performance improvements in complex urban street-canyon environments. These results confirm that the model effectively captures the combined effects of multiple diffraction, scattering, and reflection mechanisms in urban environments, exhibiting strong generalization ability and practical applicability.
\begin{table}[t]
\centering
\caption{Path Loss RMSE Comparison}
\label{tab:RMSE_results}
\begin{tabular}{l p{1.2cm} p{1.2cm} p{2.5cm}}
\toprule
\textbf{Scenario} & \textbf{Proposed Model (dB)} & \textbf{3GPP TR 37.885 (dB)} & \textbf{Model without recursive electric fields (dB)} \\
\midrule
LOS & 4.65 & 6.32 & 5.72 \\
NLOS & 3.59 & 10.69 & 6.77 \\
\bottomrule
\end{tabular}
\end{table}
\subsection{Doppler Spread Validation}
Two test routes, as illustrated in Fig. 6, are selected for comparative analysis of Doppler spread characteristics. Each route includes four sets of data: measured results, predictions from the proposed model, results from a simplified model, and estimates from the 3GPP TR 37.885 standard model. The measured Doppler spread is extracted by performing an autocorrelation operation on the channel impulse response (CIR) followed by a Fourier transform. The Doppler spread predicted by the proposed model is obtained using (21), through the calculation of Doppler shifts corresponding to each propagation path. The simplified model neglects multiple diffraction effects and retains only two dominant paths—those formed between the TX and the primary scatterers, namely the buildings located on both sides of the road closest to the receiver. The 3GPP model, based on standardized Urban Macro (UMa) scenario parameters, estimates the Doppler spread statistically using an empirical relationship between the angular spread and the maximum Doppler shift, without relying on specific geometric configurations. Its mathematical formulation can be expressed as:
\begin{equation}
\sigma_d=f_{d, \max } \cdot \sigma_\gamma
\end{equation}
Here, $\sigma_{\gamma} = 11^{\circ}$ denotes the empirical parameter of angular spread, as defined in the 3GPP channel model \cite{zhu20213gpp}. To comprehensively illustrate the performance of different models in the Doppler domain, all results are comparatively analyzed in two forms: scatter density maps and cumulative distribution functions (CDFs), as shown in Fig. \ref{fig:LOS} and \ref{fig:NLOS}. The former depicts the spatial distribution characteristics of Doppler spread, while the latter reveals the overall statistical behavior and the consistency between model predictions and measurements. In addition, we use the Kolmogorov-Smirnov-test (KS-test) to measure the distance ($D_{\text{ks}}$) between the distributions of the RMS Doppler spread of the synthetic data and measurement data, as shown in Table III. A smaller $D_{\text{ks}}$ indicates more similarities between the two distributions.
\begin{table}[t]
\centering
\caption{Doppler spread CDF $D_{\text{ks}}$ Comparison}
\label{tab:KS_results}
\begin{tabular}{l p{1.2cm} p{1.2cm} p{2.5cm}}
\toprule
\textbf{Scenario} & \textbf{Proposed Model} & \textbf{3GPP TR 37.885} & \textbf{Model without recursive electric fields} \\
\midrule
LOS & 0.09 & 0.95 & 0.33 \\
NLOS & 0.07 & 0.99 & 0.89 \\
\bottomrule
\end{tabular}
\end{table}

Overall, the Doppler spread along the LOS route is generally larger than that along the NLOS route. This difference aligns with the physical principles of electromagnetic wave propagation. In LOS scenarios, a dominant direct signal component exists, and the projection angle between the receiver’s motion direction and the incident wave direction varies significantly with position, resulting in a wider distribution of Doppler frequency shifts. In contrast, in NLOS scenarios, the direct path is completely obstructed, and signal propagation primarily relies on diffraction and reflection. Consequently, the received energy tends to be more isotropically distributed, leading to smaller overall Doppler shifts and a narrower spread range. Moreover, in heavily obstructed urban environments, the number of effective propagation paths is limited and the geometric relationships among them remain relatively stable, causing the instantaneous Doppler distribution to become more concentrated.

From the comparison, the proposed model shows good agreement with the measured data under both LOS and NLOS conditions. This consistency arises because the model explicitly accounts for key propagation mechanisms such as geometric constraints imposed by buildings and multiple diffraction effects, thereby accurately capturing the Doppler spread characteristics induced by the combined influence of diffraction, scattering, and reflection. As a result, it provides a more precise representation of the time-varying behavior of the signal. In contrast, the simplified model considers only the direct path and a single dominant scatterer, leading to Doppler spreads that are significantly lower than the measured values. The 3GPP model, which employs a fixed angular spread parameter for empirical estimation, lacks scenario-specific adaptability; consequently, its predicted Doppler spreads remain nearly identical across different conditions and deviate considerably from the experimental results.
\section{CONCLUSION}
In this paper, we propose a geometry map-based site-specific propagation model which directly extracts key geometric parameters from an urban 3D map database and, by incorporating the UTD, derives a recursive formulation for the calculation of multiple diffraction fields, enabling the derivation of essential channel characteristics such as path loss, multipath delay, and Doppler distributions. In addition, a identification algorithm is developed to efficiently filter out building structures that have a significant impact on propagation paths. The model is validated based on realistic channel measurements. The results show that in typical urban environments, the proposed model accurately characterizes path loss variations under both LOS and NLOS conditions. Particularly in NLOS scenarios, the model achieves markedly higher prediction accuracy than the 3GPP model and simplified models neglecting multiple diffraction effects, reducing RMSE by approximately 7.1 dB and 3.18 dB, respectively. Furthermore, Doppler spread analysis confirms that the model precisely characterizes time-varying propagation behaviors, achieving the closest agreement with measurements among all the compared models, thereby validating scalability and generalization of the proposed model. 

\appendix
\renewcommand{\theequation}{A.\arabic{equation}}
\setcounter{equation}{0}

The following auxiliary functions are introduced to support the geometric identification of buildings relevant to the radio propagation process. Their objective is to determine the visibility relationship between the TX and RX and to characterize the spatial positioning of surrounding scatterers in a three-dimensional environment.

The projection function $f_{\text{proj}}(p,L)$ computes the orthogonal projection of an arbitrary point $p$ onto a given line segment $L = \overline{ab}$, as expressed by
\begin{align}
f_{\text{proj}}(p,L) &= a + \frac{(p-a)\cdot (b-a)}{\lVert b-a \rVert^2}(b-a),
\quad L = \overline{ab},
\end{align}
which provides a geometric reference for assessing the relative position of buildings with respect to the TX–RX line.

The side classification function $f_{\text{side}}(p,L)$ identifies whether a point $p$ lies on the left or right side of the propagation path $L$, defined as
\begin{align}
f_{\text{side}}(p,L) &= \operatorname{sign}\!\big( (b-a) \times (p-a) \big)_z,
\quad L = \overline{ab}.
\end{align}
This function enables the algorithm to partition candidate buildings into distinct spatial regions, facilitating side-based visibility filtering.

Finally, the occlusion test function $f_{\text{block}}(a,b,B)$ determines whether the propagation segment $\overline{ab}$ is obstructed by any face of a given building $B$, formulated as
\begin{align}
f_{\text{block}}(a,b,B) &=
\begin{cases}
1, & \exists \; \text{face } F \in B \;\; \text{s.t. } \overline{ab} \cap F \neq \emptyset, \\
0, & \text{otherwise}.
\end{cases}
\end{align}
A value of 1 indicates that building $B$ blocks the direct propagation path, marking the link as NLOS. Together, these functions provide a systematic and physically interpretable framework for identifying relevant buildings and filtering scatterers in urban channel modeling.

\bibliographystyle{IEEEtran}
\bibliography{text}

\ifCLASSOPTIONcaptionsoff
  \newpage
\fi

 \end{document}